\documentclass[reprint,aps,pra,nofootinbib]{revtex4-2}

\usepackage{graphicx}
\usepackage{amsmath,amssymb}
\usepackage{epstopdf}
\usepackage{epstopdf}
\usepackage{float}
\usepackage{textcomp}
\usepackage{siunitx}
\newcommand{\ket}[1]{|#1\rangle}

\begin{document}

%
%
\title[EIT to ATS transition]{Coherence as an indicator to discern electromagnetically induced transparency and Autler-Townes splitting}
\author{Arif Warsi Laskar,$^{*,\ddagger}$ Pratik Adhikary, Niharika Singh and Saikat Ghosh$^\dagger$}
\affiliation{Department of Physics, Indian Institute of Technology, Kanpur-208016, India}

\def\thefootnote{$*$}\footnotetext{laskar@iap.uni-bonn.de}
\def\thefootnote{$\dagger$}\footnotetext{gsaikat@iitk.ac.in}
\def\thefootnote{$\ddagger$}\footnotetext{Present address: Institute of Applied Physics, University of Bonn, 53115 Bonn, Germany}
\date{\today}

\begin{abstract} 
Electromagnetically induced transparency (EIT) and Autler-Townes splitting (ATS) are generally characterized and distinguished by the width of the transparency created in the absorption profile of a weak probe in presence of a strong control field. This often leads to ambiguities, as both phenomena yield similar spectroscopic signature. However, an objective method based on Akaike's Information Criterion (AIC) test offers a quantitative way to discern the two regimes when applied on the probe absorption profile. The obtained transition value of control field strength was found to be higher than the value given by pole analysis of the corresponding off-diagonal density matrix element. By contrast, we apply the test on ground state coherence and the measured coherence quantifier, which yields a distinct transition point around the predicted value even in presence of noise. Our test accurately captures the transition between two regimes, indicating that a proper measure of coherence is essential for making such distinctions.
\end{abstract}

\maketitle

\section{Introduction}

In general, the formation of coherent superposition in atomic media due to electromagnetically induced transparency (EIT) is characterized by the narrowness of the transparency window created by a control field in the probe absorption profile~\cite{Fleischhauer05}. However, it is also well accepted that emergence of such a transparency window is not an exclusive signature for the generation of superposed states, and it can also occur due to hybridization of ground state with the excited state in presence of a strong control field usually known as Autler-Townes splitting (ATS)~\cite{Autler55,Olga08,agarwal10,Salloum10,sanders11}. In particular, for systems with comparatively larger decoherence rates such as room-temperature atoms~\cite{arif16,arif18}, semiconductors~\cite{Brunner70}, rare-earth doped materials~\cite{Hemmer01,Manson05}, photonic crystals~\cite{Wei09}, plasmonic systems~\cite{Xiang08} and quantum dots~\cite{Hopkinson09}, rates of classical dynamics tend to be comparable or even dominate over quantum time scales. In such systems, often high control powers are needed to observe any signature of EIT. However, such strong control strengths ($>$ excited state decay rate) can result in an ATS. 

Although in ATS regime, there is an absence of effective ground state superposition, this phenomenon is of significance in several spectroscopic applications~\cite{Zhang14}. Furthermore, it has been used for measuring transition dipole moment~\cite{Yakshina11}, quantum control of spin-orbit interaction~\cite{Ahmed11}, storing photon~\cite{Saglamyurek2018,Rastogi21}, and angular momentum alignment of non-polar molecules~\cite{Qi99}. On the other hand, the regime of EIT has been a major workhorse for creating a stable superposition of ground states and therefore has been widely used in quantum technologies, stopping and storing light~\cite{Lukin01,Bajcsy03,Simon07,Choi08,Lvovsky09,Tanji09,Heinze13,Kuzmich13}, leading to the realization of quantum synchronization~\cite{Arif20}, and enhancing optical non-linearity at the single photon level~\cite{Tanji-Suzuki1266,Vuletic12,Firstenberg13}. In particular, with growing interest of using coherently prepared atomic media for quantum technologies, it becomes crucial to distinguish EIT and ATS regimes and quantify effective, long lived ground state superposition in a physical system.

Despite differing underlying physics, the two regimes yield similar spectroscopic signatures, leading to ambiguity in differentiating the two processes from a direct comparison of the absorption profiles. Anisimov \textit{et al.}~\cite{sanders11} proposed a method based on Akaike's information criterion (AIC)~\cite{Burnham02} test to objectively discern the two regimes. In particular, they applied the test on absorption profiles proportional to the off-diagonal density matrix element $\mathrm{Im}(\rho_{13})$ of a weak probe, with increasing control field strength $\Omega_\mathrm{c}$. They observed that the transition between the two regions occurred  at $\Omega_\mathrm{c}^\mathrm{t}/\gamma_{13}=0.86$. When a per-point AIC test was carried out to incorporate the experimental noise, the transition from EIT to ATS occurred smoothly over a certain region, without a signature of sharp transition. Based on this proposed method to quantitatively distinguish EIT from ATS regime, several studies have focused on discriminating the two phenomena in different  systems such as atomic system~\cite{Laurat13,Liu20,Ji21}, micro-resonators~\cite{Peng2014,He15}, superconducting circuits~\cite{Liu16}, mechanical resonator~\cite{Yi19}, and photonic crystal ~\cite{Tian22}. Although such tests provides a quantitative indicator of the transition point between the two regions, the obtained transition point is larger than the theoretically predicted value estimated from the pole analysis of $\rho_{13}$ which is discrete at $\Omega_\mathrm{c}^\mathrm{t}/\gamma_{13}=0.5$~\cite{Olga08,Salloum10}. Therefore, there is a discrepancy between the experimental and theoretical results.

We believe that such discrepancy in theory and experiment occurs because the AIC test is applied directly on probe absorption [$\propto\mathrm{Im}(\rho_{13})$], which shows a similar signature in both regimes. On the contrary, in Ref.~\cite{arif18}, we observed that the ground state coherence or $|\rho_{12}|$ has spectroscopically distinct shapes in the two regimes. This suggests an alternative way of directly applying AIC test on the ground state coherence to distinguish the two regimes unambiguously. Here, we explore this possibility.

\begin{figure}[htbp]
	\centering\includegraphics[width=\linewidth]{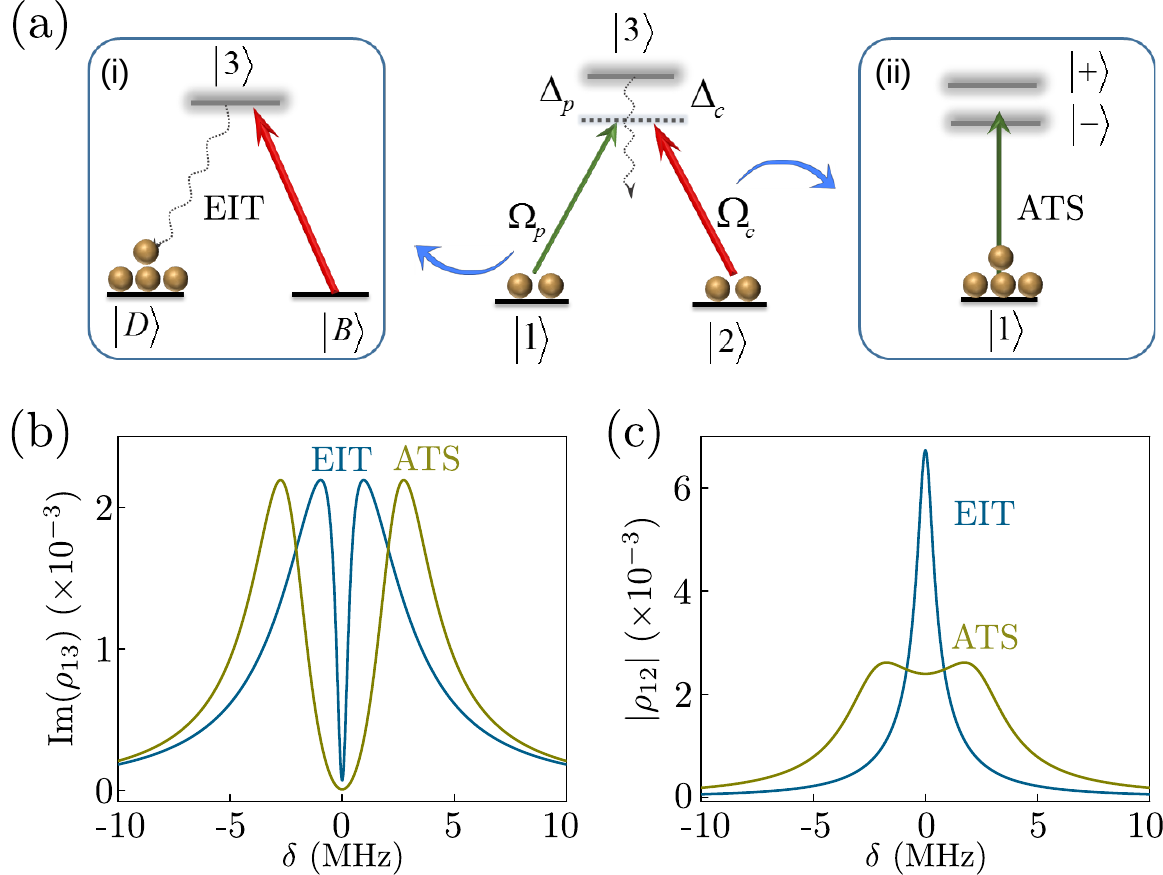}
	\caption{(a) Equivalent atomic energy level diagram for EIT (left) and ATS (right) regime in a three-level atomic system. (b) Typical probe response, and blue and dark yellow curves represent the EIT and ATS cases, respectively. (c) Absolute value of $\rho_{12}$, and blue and dark yellow curves are for EIT and ATS case, respectively.}
	\label{fig:EIT_ATS} 
\end{figure}

Figure~\ref{fig:EIT_ATS}(a) shows an equivalent atomic energy level schemes of EIT [Fig.~\ref{fig:EIT_ATS}(a), left] and higher field ATS [Fig.~\ref{fig:EIT_ATS}(a), right] regimes in a three-level atomic system driven by a weak probe ($\Omega_\mathrm{p}$) and a control field ($\Omega_\mathrm{c}$). When the control field is relatively weak, the atoms are driven into a superposition of ground states ($\ket{1}$ and $\ket{2}$) or dark state $\ket{D}$ [Fig.~\ref{fig:EIT_ATS}(a), left], leading to a sharp transparency window in the absorption profile of a weak probe field [Fig.~\ref{fig:EIT_ATS}(b)]. However, a higher control field hybridizes the ground state $\ket{2}$ with the excited state $\ket{3}$ to fragile states $\ket{+}$ and $\ket{-}$ [Fig.~\ref{fig:EIT_ATS}(a), right], which creates a doublet structure in probe absorption or ATS [Fig.~\ref{fig:EIT_ATS}(b)].

In particular, we compare the transition between EIT and ATS regimes for three-level systems based on the two methods: from the shape of the probe absorption profile (Im($\rho_{13}(\delta)$)), and from the ground state coherence ($|\rho_{12}(\delta)|$). Figure~\ref{fig:EIT_ATS}(b) shows Im($\rho_{13}$) as a function of two-photon detuning ($\delta$) for both the EIT and ATS cases, depicting a splitting in both scenarios. On the contrary, $|\rho_{12}|$ shows completely distinct shapes: a single resonance peak in EIT regime and a splitting in ATS regime [Fig.~\ref{fig:EIT_ATS}(c)]. A mere visual inspection can resolves the ambiguity to distinguish between the two regimes. To quantitatively compare the advantage of using coherence, we  apply the AIC test to these shapes, i.e., on the ground state coherence ($|\rho_{12}|$) and compare it with the corresponding probe absorption profile [$\propto$ Im($\rho_{13}$)]. Here, we show that the test on EIT and ATS for $|\rho_{12}|$ clearly identifies the two regimes even in presence of noise, while the test on Im$(\rho_{13})$ fails to do so. The AIC test indeed predicts a transition point at $\Omega_\mathrm{c}^\mathrm{t}/\gamma_{13}=0.5$, consistent with the existing predictions~\cite{Olga08,Salloum10}. Furthermore, we apply the test on experimentally measured coherence quantifier $C$ of  ground state coherence ($|\rho_{12}|$) which is based on a single shot time domain measurement~\cite{arif18}. In accordance with theory, the test on quantifier indeed demonstrates that a transition occurs at $\Omega_\mathrm{c}^\mathrm{t}/\gamma_{13}\simeq0.45$, which is very close to the predicted value. This small deviation in the transition point for quantifier $C$ from the numerically computed $|\rho_{12}|$ can be understood by taking into account the finite ground state decoherence ($\gamma_{12}$).

\section{Theoretical analysis of EIT-ATS model}

In order to apply the AIC test on shapes of $|\rho_{12}|$ and Im($\rho_{13}$), we first derive the expressions for both quantities in EIT and ATS regimes for a three-level atomic system [Fig. \ref{fig:EIT_ATS}(a)] using the similar approach as in Ref.~\cite{Olga08}. Under semi-classical description of laser-atom interaction, the steady state solutions of $\rho_{12}$ and $\rho_{13}$ are~\cite{Olga08,arif18}
\begin{equation}
	\rho^{\mathrm{ss}}_{12}= \frac{\Omega_\mathrm{p}^*\Omega_\mathrm{c}/(\delta-i\gamma_{12})}{\delta+\Delta_\mathrm{c}-i\gamma_{13}-|\Omega_\mathrm{c}|^2/(\delta-i\gamma_{12})}, 
	\label{eqn:rho12}
\end{equation}

and 

\begin{equation}
	\rho^{\mathrm{ss}}_{13}= \frac{-\Omega_\mathrm{p}^*}{\delta+\Delta_\mathrm{c}-i\gamma_{13}-|\Omega_\mathrm{c}|^2/(\delta-i\gamma_{12})}, 
	\label{eqn:rho13}
\end{equation}
where $\delta=\Delta_\mathrm{p}-\Delta_\mathrm{c}$, $\Delta_\mathrm{p}$, $\Delta_\mathrm{c}$, $\gamma_{13}$, $\gamma_{12}$, $\Omega_\mathrm{c}$, and $\Omega_\mathrm{p}$ correspond to two-photon detuning, probe detuning, control detuning, excited state decay rate from $\ket{3}$ to $\ket{1}$, ground state decoherence, control Rabi frequency, and probe Rabi frequency, respectively. The dressed state energies and widths can be obtained by the poles of Eqs.~\ref{eqn:rho12} and \ref{eqn:rho13} as
\begin{widetext}
	\begin{equation}
		\delta_{\pm}=\frac{1}{2}\left[-\Delta_\mathrm{c}+i(\gamma_{13}+\gamma_{12})\pm\sqrt{4|\Omega_\mathrm{c}|^2+(\Delta_\mathrm{c}-i(\gamma_{13}-\gamma_{12}))^2}\right].
	\end{equation}
\end{widetext}
The two poles give the two resonant contributions to the atomic response, and  Re$(\delta_\pm)$ and Im$(\delta_\pm)$ correspond to the frequency and dephasing rate of the dressed states, respectively. Accordingly, $\rho_{12}^\mathrm{ss}$ and $\rho_{13}^\mathrm{ss}$ can be expressed as a superposition of these resonant contributions~\cite{Olga08}:
\begin{equation}
	\rho^{\mathrm{ss}}_{12}=\frac{\Omega^*_\mathrm{p}B_+}{\delta-\delta_+}+\frac{\Omega^*_\mathrm{p}B_-}{\delta-\delta_-},
\end{equation}
and
\begin{equation}
	\rho^{\mathrm{ss}}_{13}=\frac{\Omega^*_\mathrm{p}A_+}{\delta-\delta_+}+\frac{\Omega^*_\mathrm{p}A_-}{\delta-\delta_-}.
\end{equation}
where $B_{\pm}=\pm\Omega_\mathrm{c}/(\delta_+-\delta_-)$ and $A_{\pm}=\pm(\delta_{\pm}-i\gamma_{12})/(\delta_+-\delta_-)$ are the strengths of two contributions for $\rho_{12}^\mathrm{ss}$ and $\rho_{13}^\mathrm{ss}$, respectively.

\begin{figure}[htbp]
	\centering\includegraphics[width=\linewidth]{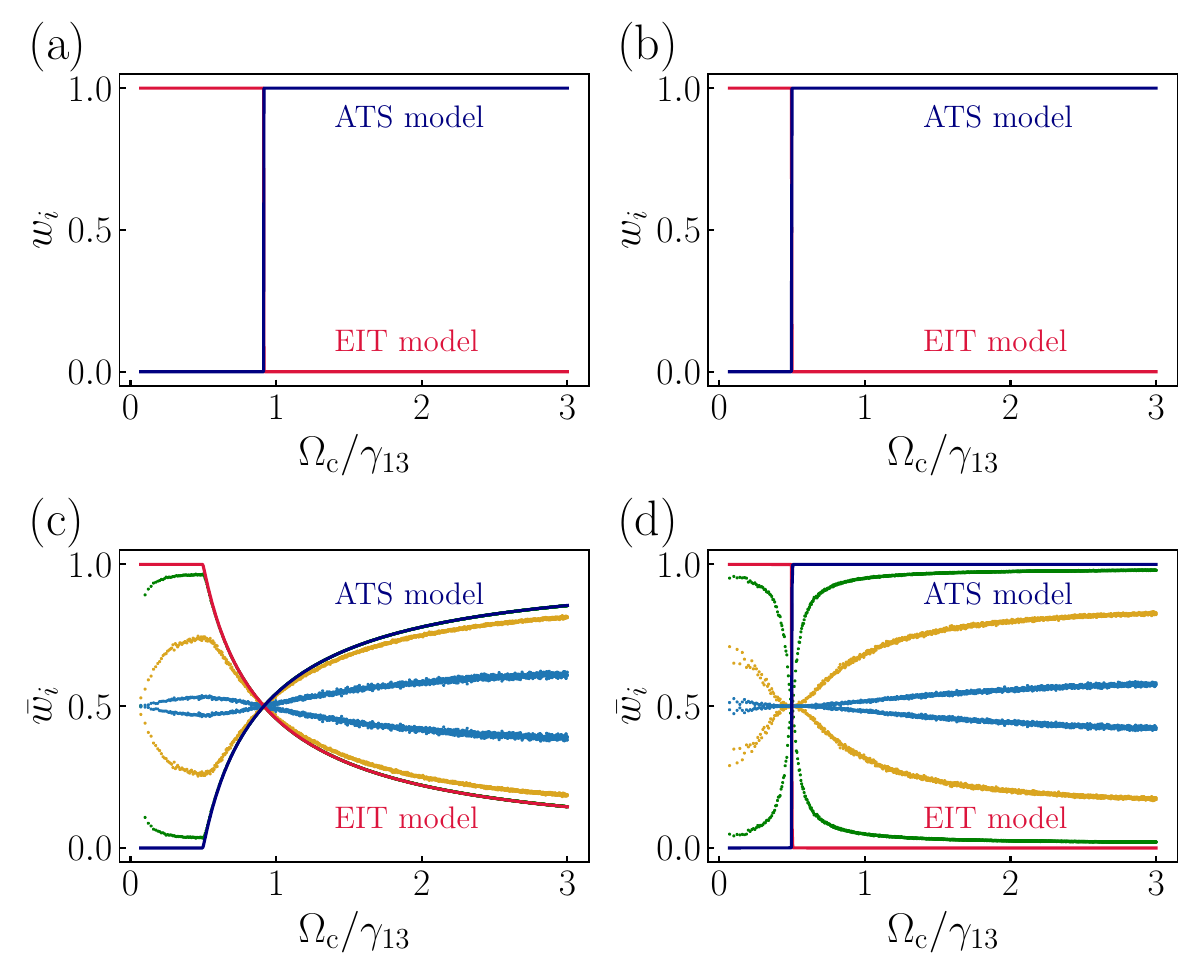}
	\caption{(a) and (b) AIC weight comparison with control field strength ($\Omega_\mathrm{c}/\gamma_{13}$) of models for Im($\rho_{13}$) ($A_{i}$) and  $|\rho_{12}|$ ($B_{i}$), respectively, where $i$ represents the EIT (red line) and ATS (blue line) models. The transition point obtained from shape comparison of Im($\rho_{13}$) is $\Omega_\mathrm{c}^\mathrm{t}/\gamma_{13}=0.91$ while $|\rho_{12}|$ gives the transition at theoretically predicted value $\Omega_\mathrm{c}^\mathrm{t}/\gamma_{13}=0.5$. (c), (d) per point Akaike weight of $A_i$ (c) and $B_i$ (d) models with control field strength. (c) shows distinct three region, (i) $\Omega_\mathrm{c}/\gamma_{13}<0.5$ where EIT model (red line) dominates unconditionally , (ii) $0.5<\Omega_\mathrm{c}/\gamma_{13}<0.91$ ATS model (navy-blue line) start dominating and the system goes through an inconclusive region, (iii) $\Omega_\mathrm{c}/\gamma_{13}\geq0.91$ ATS start to dominate. With introduction of Gaussian ($\sigma=0.01$ (green dots), $\sigma=0.1$ (yellow dots) and $\sigma=0.5$ (light-blue dots)) the unconditional domination of EIT model for $\Omega_\mathrm{c}/\gamma_{13}<0.5$ does not hold any more. While (d) shows two distinct regions. For $\Omega_\mathrm{c}/\gamma_{13}<0.5$ EIT dominate unconditionally (red line) whereas ATS (navy-blue line) model surpass for $\Omega_\mathrm{c}/\gamma_{13}\geq0.5$. Even after introduction of Gaussian noise ($\sigma=0.01$ (green dots), $\sigma=0.1$ (yellow dots) and $\sigma=0.5$ (light-blue dots)) the two regions are clearly separated out at $\Omega_\mathrm{c}/\gamma_{13}=0.5$ without any ambiguity. Here, $\gamma_{13}$ and  $\gamma_{12}$ are set to 3 MHz and 0.01 MHz, respectively.}
	\label{fig:wAIC}
\end{figure}

For a resonant case $\Delta_\mathrm{c}=0$, EIT occurs at $\Omega_\mathrm{c}<(\gamma_{13}-\gamma_{12})/2$ where the two dressed state share the same reservoir similar to Fano interference, and for $\Omega_\mathrm{c}>(\gamma_{13}-\gamma_{12})/2$, the dressed states decay into distinct reservoirs, which give rise to two resonant contributions corresponding to ATS. In EIT region, it can be seen that $\mathrm{Re}(\delta_\pm)=0=\mathrm{Re}(B_{\pm})$, which implies that the two dressed states coincide at $\delta=0$ with different dephasing rates given by Im$(\delta_{\pm})$. 
Accordingly, the profile of $|\rho_{12}|$ in EIT regime can be approximated to $B_{\text{EIT}}(\delta)=P_1\left[(\gamma_{+}-\gamma_{-})/\sqrt{(\delta^2+\gamma_{+}^2)(\delta^2+\gamma_{-}^2)}\right]$. For high control field where $B_\pm\approx1/2$ and $\delta_{\pm}\approx\pm\Omega_\mathrm{c}+i\gamma_{13}/2$, $|\rho_{12}|$ can be approximated as $B_{\text{ATS}}(\delta)=P_2\left[|\Omega|/\sqrt{\delta^4+2\delta^2(\gamma^{2}-|\Omega|^2)+(\gamma^{2}+|\Omega|^2)^2}\right]$. While $B_{\text{EIT}}(\delta)$ has a single peak structure, $B_{\text{ATS}}(\delta)$ shows two distinct peaks.

Similar pole analysis for Im($\rho_{13}$) yields distinct functions for EIT and ATS regions~\cite{sanders11}, which are $A_\text{EIT}(\delta)=S_+^2/(\delta^2+\gamma_{+}^2)-S_-^2/(\delta^2+\gamma_{-}^2)$ and  $A_\text{ATS}(\delta)=S^2[1/((\delta-\delta_0)^2+\gamma^2)+1/((\delta+\delta_0)^2+\gamma^2)]$, respectively. Both $A_\text{EIT}(\delta)$ and $A_\text{ATS}(\delta)$ have a two peak structure. This leads to a lack of clear distinction between the two regimes as compared to $|\rho_{12}|$.

\section{Model based comparison of EIT-ATS using AIC}
As proposed in Ref.~\cite{sanders11}, the AIC test~\cite{Burnham02} allows one to select the model with smallest AIC value as the best model while comparing the shape of EIT and ATS. We obtained the AIC value directly from the model fit, which is given by $I_i=-2\text{log}L_i+2K_i$, where $L_i$ is the maximum likelihood and $K_i$ is the number of parameter used to fit the $i^\mathrm{th}$ model, and $i$ denotes EIT or ATS. To compare EIT and ATS models, Anisimov et al.~\cite{sanders11} used the Akaike weight which gives the relative likelihood of $i^\mathrm{th}$ model and given by $w_i=e^{-\Delta I_i/2}/\sum_{m=1}^{2}e^{-\Delta I_m/2}$, where $\Delta I_i$ is the relative AIC for $i^\mathrm{th}$ model with respect to minimum AIC value from the set of models. We apply the AIC test by fitting $A_i(\delta)$ and $B_i(\delta)$ models to numerically computed $\mathrm{Im}(\rho_{13}(\delta))$ ($\propto$ probe absorption profile) and coherence profile $|\rho_{12}(\delta)|$, respectively for a three level atomic system (Fig.\ref{fig:EIT_ATS}). These fits are performed across a range of control field strengths, and for the fitting process, we utilize the nonlinear regression model fit function available in Matlab, which directly provides AIC value ($I_i$). The relative likelihood of $i^\mathrm{th}$ model is then evaluated by calculating Akaike weight $w_i$, for different control field strength ($\Omega_\mathrm{c}/\gamma_{13}$), as shown in Figs.~\ref{fig:wAIC}(a) and (b) corresponding to the $A_i$ and $B_i$ models, respectively. As the relative likelihood describing the Akaike weight $w_i$ of these models can only take 0 or 1, it conclusively determine the transition from  EIT (red line) to ATS (navy-blue line) regimes. However, the AIC test based on $A_i$ model (Im($\rho_{13}$)) [Fig.~\ref{fig:wAIC}(a)]  yields higher transition point at $\Omega_\mathrm{c}^\mathrm{t}/\gamma_{13}=0.91$, while the test based on $B_i$ ($|\rho_{12}|$) [Fig.~\ref{fig:wAIC}(b)] accurately captures the transition at the theoretically predicted value $\Omega_\mathrm{c}^\mathrm{t}/\gamma_{13}=0.5$~\cite{Olga08}. This considerable discrepancy in the comparison between these two models can be understood by examine the shape of  Im($\rho_{13}$) and $|\rho_{12}|$ with two-photon detuning ($\delta$) [Figs.~\ref{fig:EIT_ATS}(b) and (c)]. The probe absorption (Im($\rho_{13}$)) exhibits similar features in both the EIT and ATS regimes (two peak structure), causing both the $A_\mathrm{EIT}$ and $A_\mathrm{ATS}$ models to fit effectively near the transition point. The test prefers $A_\mathrm{EIT}$ model over $A_\mathrm{ATS}$ model beyond the actual transition point up to a certain control field strength, here in our test, it occurs at $\Omega_\mathrm{c}^\mathrm{t}/\gamma_{13}=0.91$. In contrast, coherence profile $|\rho_{12}|$ exhibits distinct features (a single peak and a double peak structure in EIT and ATS regimes, respectively), allowing the models  ($B_\mathrm{EIT}$ and $B_\mathrm{ATS}$) to fit more accurately in their respective regime and reveal the actual transition.

\begin{figure}[htbp]
	\centering\includegraphics[width=\linewidth]{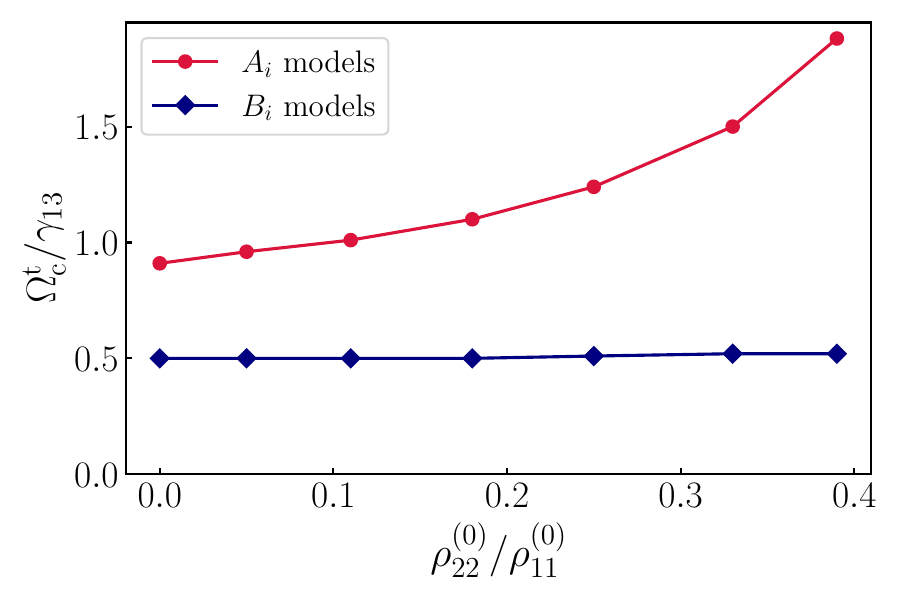}
	\caption{The transition point $\Omega_\mathrm{c}^\mathrm{t}/\gamma_{13}$ between EIT and ATS regimes obtained from Akaike weight ($w_i$)  comparison of $A_i$ (red dots) and $B_i$ (blue diamond) models for a system where incoherent effects such as optical pumping redistribute the population. Here, $\rho_{11}^{(0)}$ and $\rho_{22}^{(0)}$ are the steady state population in probe and control field ground state, respectively.}
	\label{fig:Transition}
\end{figure} 

For experimental data, where various noise channels make the selection of models difficult, the model comparison is done by per-point (mean) AIC weight $\bar{w}_i=e^{-\Delta \bar{I}_i/2}/\sum_{m=1}^{2}e^{-\Delta \bar{I}_m/2}$, where $\bar{I}_i=I_i/n$ and n is the number of data points. Figures~\ref{fig:wAIC}(c) and (d) show the comparison of $\bar{w}_i$ for $A_i$ and $B_i$ models, respectively. Figure~\ref{fig:wAIC}(c) shows three regions when the coupling strength is increased gradually in the absence of noise (red line: EIT model and navy-blue line: ATS model). In region I ($\Omega_\mathrm{c}/\gamma_{13}<0.5$), the EIT model dominates unconditionally. However, in region II ($0.5<\Omega_\mathrm{c}/\gamma_{13}<0.91$) ATS model starts to have non zero value, where none of the models can be ruled out. When the control field strength is further increased ($\Omega_\mathrm{c}/\gamma_{13}>0.91$) ATS model starts to dominate over EIT model. With introduction of Gaussian noise ($\sigma=0.01$ (green dots), $\sigma=0.1$ (yellow dots) and $\sigma=0.5$ (light-blue dots)), the unconditional domination of EIT model for $\Omega_\mathrm{c}/\gamma_{13}<0.5$ does not hold anymore [Fig.~\ref{fig:wAIC}(c)]. However, the model comparison of $B_i$ shows a sharp transition at $\Omega_\mathrm{c}^\mathrm{t}/\gamma_{13}=0.5$ [Fig.~\ref{fig:wAIC}(d)]. Below and above this transition value, EIT (red line) and  ATS models (navy-blue line) dominate unconditionally, respectively. It shows two distinct regions even in presence of Gaussian noise ($\sigma=0.01$ (green dots), $\sigma=0.1$ (yellow dots) and $\sigma=0.5$ (light-blue dots)).

\section{Optical pumping effects on model comparison}
 Moreover, in systems that are relatively dominated by various incoherent effects, such as room temperature atoms, where incoherent processes compete with its quantum counterpart~\cite{arif16}, the steady state probe response (Im($\rho_{13}$)) settle at a different level due to optical pumping, losses, and thermal diffusion of atoms compared to the ideal EIT case (Eq.~\ref{eqn:rho13}). However, the ground state coherence $|\rho_{12}|$ remains unaffected by these incoherent effects. In such scenarios, the test based on $|\rho_{12}|$ ($B_i$) can outperform the test which are relied on Im($\rho_{13}$) ($A_i$).
 
We apply the AIC test for such systems, focusing on $\rho_{12}^\mathrm{ss}$ and $\rho_{13}^\mathrm{ss}$ while taking into account the contributions from the resulting steady state population distribution $\rho_{ii}^{(0)}$ for i=1, 2 and 3 levels. The expression for $\rho_{12}^\mathrm{ss}$ and $\rho_{13}^\mathrm{ss}$ corrected to first order, can be written as follows:
\begin{widetext}
\begin{equation}
	\rho^{\mathrm{ss}}_{12}=\frac{\Omega_\mathrm{p}^*\Omega_\mathrm{c}/\left(\delta-i\gamma_{12}\right)}{\delta+\Delta_\mathrm{c}-i\gamma_{13}-|\Omega_\mathrm{c}|^2/\left(\delta-i\gamma_{12}\right) }\left[\left(\rho_{11}^{(0)}-\rho_{33}^{(0)}\right)-\frac{\delta+\Delta_\mathrm{c}-i\gamma_{13}}{\Delta_\mathrm{c}+i\gamma_{23}}\left(\rho_{22}^{(0)}-\rho_{33}^{(0)}\right)\right],\nonumber
	\label{eqn:rho12ss}
\end{equation}
and 
\begin{equation}
	\rho^{\mathrm{ss}}_{13}= \frac{-\Omega_\mathrm{p}^*}{\delta+\Delta_\mathrm{c}-i\gamma_{13}-|\Omega_\mathrm{c}|^2/(\delta-i\gamma_{12})}\left[\left(\rho_{11}^{(0)}-\rho_{33}^{(0)}\right)-\frac{|\Omega_\mathrm{c}|^2}{(\Delta_\mathrm{c}+i\gamma_{23})(\delta-i\gamma_{12})}\left(\rho_{22}^{(0)}-\rho_{33}^{(0)}\right)\right], \nonumber
	\label{eqn:rho13ss}
\end{equation}
\end{widetext}
where $\gamma_{23}$ is the decay rate from $\ket{3}$ to $\ket{2}$, $\delta$, $\Delta_\mathrm{p}$, $\Delta_\mathrm{c}$, $\gamma_{13}$, $\gamma_{12}$, $\Omega_\mathrm{c}$, and $\Omega_\mathrm{p}$ are defined as previously. We fit $A_i(\delta)$ and $B_i(\delta)$ models to the numerically computed profiles of Im($\rho_{13}^\mathrm{ss}(\delta)$) and $|\rho_{12}^\mathrm{ss}(\delta)|$, respectively and calculate Akaike weight ($w_i$) for both models while varying control field strength, as described in section 3. The obtained transition point ($\Omega_\mathrm{c}^\mathrm{t}/\gamma_{13}$) from these models comparison is shown in Fig.~\ref{fig:Transition} for the respective models ($A_i$, red dot and $B_i$, blue diamonds) with varying ground state population distribution ($\rho_{22}^{(0)}/\rho_{11}^{(0)}$). Here, $\rho_{33}^{(0)}$ is set to 0 for simplicity. We observed that with increasing population in control field ground state ($\rho_{22}^{(0)}$) the transition point for $A_i$ model shifts to the higher values, while for $B_i$ model, the transition point remains close to $\Omega_\mathrm{c}^\mathrm{t}/\gamma_{13}=0.5$. This observation shows that when the system is dominated by incoherent processes, the AIC test based on coherence ($|\rho_{12}|$) can outperform the comparison based on probe absorption (Im($\rho_{13}$)) in capturing the true transition from EIT to ATS regime.

\begin{figure}[htbp]
	\centering\includegraphics[width=\linewidth]{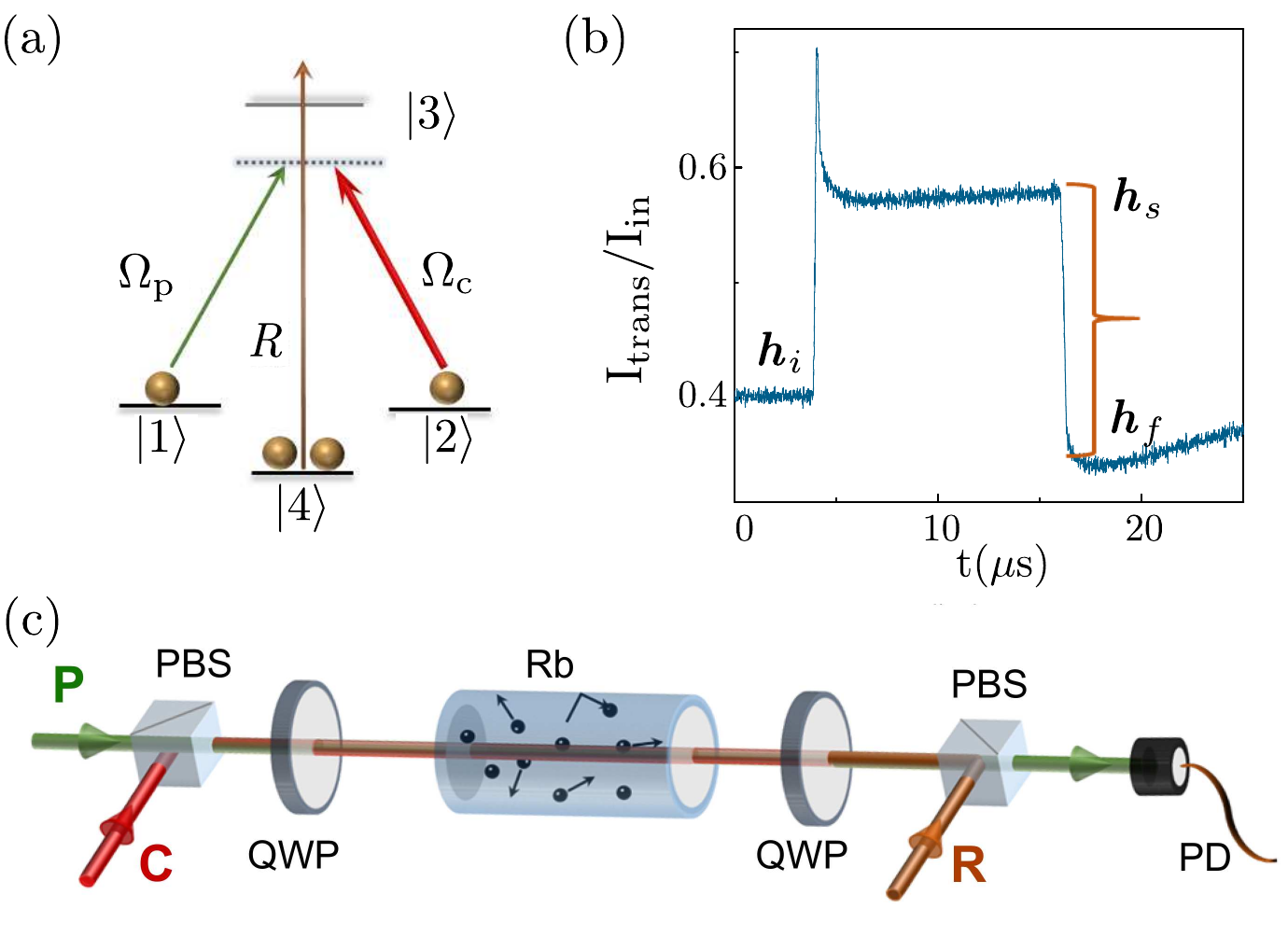}
	\caption{ (a) A three-level system is considered by connecting the ground states $\ket{1}$ and $\ket{2}$ to $\ket{3}$ with probe (Rabi frequency $\Omega_\mathrm{p}$) and a 10 $\mu$s pulsed control (Rabi frequency $\Omega_\mathrm{c}$), respectively. A repumper R is used to experimentally simulate a closed three-level atomic system. (b) Typical time response of probe transmission at $\delta=0$. (c) Experimental setup: P: Probe, C: Control, R: Repumper, PBS: Polarizing beam splitter, QWP: Quarter-waveplate, Rb: Rubidium vapor cell, PD: photo diode. Here, $\Omega_\mathrm{p}=2.5 \times 10^{-3}\gamma_3$,  $\Omega_\mathrm{c}=2.3\times 10^{-1}\gamma_3$ and the repumping rate R $=2.7\times10^{-1}\gamma_3$, where $\gamma_3= (2\pi)6.0~\mathrm{MHz}$ is the decay rate from level $\ket{3}$, and the diameter of all the laser beams is $\sim$ 4 mm.}
	\label{fig:setup}
\end{figure}

\begin{figure}[htbp]
	\centering\includegraphics[width=\linewidth]{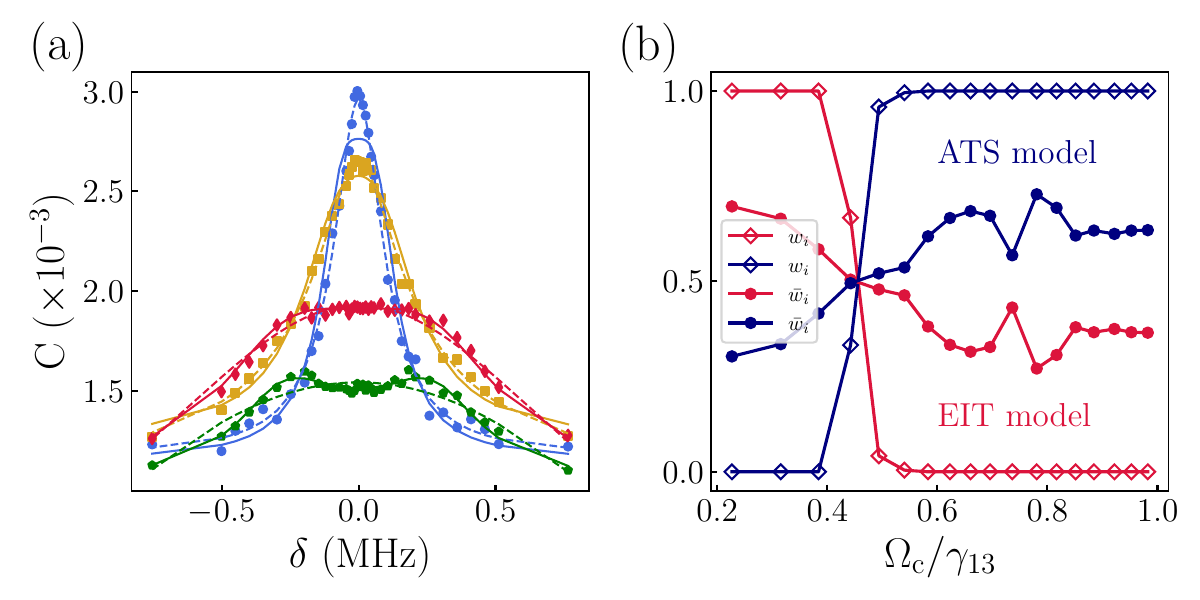}
	\caption{ (a) Measured coherence quantifier $C(\delta)$ as a function of two photon detuning $\delta$ in symbols with the best fit of $B_\mathrm{EIT}$ (dotted line) and $B_\mathrm{ATS}$ (solid line) models for varying $\Omega_\mathrm{c}/\gamma_{13}$ = 0.23 (blue circle), 0.32 (yellow square), 0.58 (red diamond), and 0.82 (green pentagon). (b) Akaike weight $w_i$ (unfilled diamond) and per-point Akaike weight $\bar{w}_i$ (filled circle) are shown with respect to $\Omega_\mathrm{c}/\gamma_{13}$ which are obtained by fitting the coherence quantifier profile $C(\delta)$ in (a) with $B_\mathrm{EIT}$ (red) and $B_\mathrm{ATS}$ (blue) models for different control field strength. The comparison yields a transition point around $\Omega_\mathrm{c}^\mathrm{t}/\gamma_{13}\simeq\mathrm{0.45}$. Here, the experimental parameter $\Omega_\mathrm{p}$ and R are same as in Fig.~\ref{fig:setup}.}
	\label{fig:eAIC}
\end{figure} 

\section{AIC comparison on coherence quantifier}
In previous sections, we demonstrated that AIC test applied to the coherence $|\rho_{12}|$ ($B_i$) serves as a robust indicator in determining the transition from EIT to ATS regime, even in systems dominated by incoherent processes. To further validate this hypothesis, we performed the AIC test for the $B_i$ model on the measured coherence quantifier ($C$) in a thermal ensemble of atoms, using the experimental method describe in Ref~\cite{arif18}. Here, we briefly introduce the method of measuring the coherence quantifier. The idea behind this measurement relies on the fact that in a thermal ensemble of atoms, the probe response is influenced by both incoherent and coherent processes, which can be effectively separated due to their significantly different time scales~\cite{arif16}. This is particularly apparent in the steady state solution of the corresponding density matrix elements for a three level atomic system  [Fig.~\ref{fig:setup}(a)] that drives the probe absorption, which is given by~\cite{arif18}
\begin{equation}
	\rho_{13}^\mathrm{ss} = \left[-i\Omega_\mathrm{c}^*\rho_{12}^\mathrm{ss}-i\Omega_\mathrm{p}^*\left(\rho_{11}^\mathrm{ss}-\rho_{33}^\mathrm{ss}\right)\right]/\Gamma_{13},\nonumber
	\label{eqn:rho13sss}
\end{equation}
where $\Gamma_{13}=i\Delta_\mathrm{p}+\gamma_{13}$ with $\Delta_p$ the probe detuning. It is evident from the equation that the probe response in steady state ($\propto\mathrm{Im}(\rho_{13})$) consist of the coherence $\rho_{12}^\mathrm{ss}$ coupled with the control field $\Omega_\mathrm{c}$ and the incoherent population $\rho_{11}^\mathrm{ss}$ and $\rho_{33}^\mathrm{ss}$ which are modified from their thermal equilibrium by optical pumping and losses. In the absence of control field, the populations equilibrate by thermal diffusion over a slow time scale of 10 $\mu$s, which is much longer than the time scales of atom dynamics ($\sim$ 30 ns). This diffusion time corresponds to the time it takes for room temperature rubidium atoms to diffuse across a 4 mm laser beam diameter. Based on this differing time scale, we extract the amount of coherence  ($|\rho_{12}^\mathrm{ss}|$) developed in the system by taking the difference of the probe response in the presence and absence of the control field and normalizing it with the initial response when the control field is not applied. This procedure effectively removes the contribution from populations dynamics. Accordingly, the coherence quantifier can be defined as
\begin{equation}
	C = \frac{\left|\rho_{13}^{\Omega_\mathrm{c}^{\mathrm{on}},\mathrm{ss}} - \rho_{13}^{\Omega_\mathrm{c}^{\mathrm{off}},\mathrm{ss}}\right|}{ \left|\rho_{13}^{\Omega_\mathrm{c}^{\mathrm{off}},\mathrm{i}}\right|}\frac{\left|\Omega_\mathrm{p}\right|}{\left|\Omega_\mathrm{c}\right|}\rho_{11}^\mathrm{i},
	\label{eqn:definedquantifier}
\end{equation}
where $\rho_{13}^{\Omega_\mathrm{c}^{\mathrm{on}},\mathrm{ss}}$, $\rho_{13}^{\Omega_\mathrm{c}^{\mathrm{off}},\mathrm{ss}}$, and $\rho_{13}^{\Omega_\mathrm{c}^{\mathrm{off}},\mathrm{i}}$ are the steady state probe response at control-on, control-off, and the initial response in the absence of control field, respectively, and $\rho^{i}_{11}$ is the initial population in the probe ground state while the initial excited state population $\rho^{i}_{33}$ is close to 0. Experimentally, we extract the coherence from the transient response of the transmitted probe intensity ($\mathrm{I}_\mathrm{trans}$) for a closed three-level system [Fig.~\ref{fig:setup}(a)] shown in Fig.~\ref{fig:setup}(b). The transmitted intensity is related to the atomic response by $\mathrm{I}_\mathrm{trans}=\mathrm{I}_\mathrm{in}\mathrm{e}^{-\alpha\mathrm{Im}(\rho_{13})}$, where $\alpha$ is a constant containing information about atomic properties, the number of atoms, length of interaction, and wavelength etc. By utilizing Eq.~\ref{eqn:definedquantifier} and the relationship between transmitted intensity ($\mathrm{I}_\mathrm{trans}$) and $\mathrm{Im}(\rho_{13})$, we derive the coherence quantifier as follows:
\begin{equation}
	C=\frac{\left|\ln(h_s)-\ln(h_f)\right|}{\left|\ln(h_i)\right|}\frac{\left|\Omega_\mathrm{p}\right|}{\left|\Omega_\mathrm{c}\right|}\rho^{i}_{11}, 
	\label{eqn:quantifier} 
\end{equation}
where $h_i,h_s$ and $h_f$ are the transmitted intensity ($\mathrm{I}_\mathrm{trans}$) normalized with input intensity ($\mathrm{I}_\mathrm{in}$) for initial level, steady state level and fall end, respectively [Fig.~\ref{fig:setup}(b)], and here the initial population ( $\rho^{i}_{11}$ ) in the probe ground state takes the value 0.5 for a thermal ensemble of atoms. 

A three-level atomic system in Fig.~\ref{fig:setup}(a) is realized by considering $\ket{1}\equiv\ket{F=2, m_{F}}$,  $\ket{2}\equiv\ket{F=2,m_{F}-2}$ and $\ket{3}\equiv\ket{F^{\prime}=1, m_{F}-1}$ within $D_{2}$ transition of $^{85}$Rb atoms. A continuous probe $\Omega_\mathrm{p}$ and a pulsed control $\Omega_\mathrm{c}$ (width 10 $\mu$s) with opposite circular polarizations are used to drive the transitions $\ket{1}\rightarrow\ket{3}$ and $\ket{2}\rightarrow\ket{3}$, respectively. These are derived from a single laser locked at 19 MHz red detuned to the transition $\ket{F=2}\rightarrow\ket{F'=1}$ and sent through a rubidium vapor cell of length 8 cm and diameter of 2 cm,  magnetically shielded by three layers of $\mu$-metal sheet. A continuous counter propagating repumper beam locked at $\ket{F=3}\rightarrow\ket{F^{\prime}=3}$ is used to experimentally simulate a closed three-level system. A schematic of the experimental setup is shown in Fig.~\ref{fig:setup}(c). The repetition time of the whole experiments is 50 $\mu$s. More details on the experiment can be found in our earlier work~\cite{arif18}. 

The coherence quantifier $C$ (Eq.~\ref{eqn:quantifier}) is evaluated using $h_i$, $h_s$, and $h_f$ from the transient response of the probe field given in Fig.~\ref{fig:setup}(b) at a particular two-photon detuning ($\delta$) and control field strength ($\Omega_\mathrm{c}/\gamma_{13}$). $C(\delta)$ is obtained by repeating the experiment by varying $\delta$ and shown in Figure~\ref{fig:eAIC}(a) for several control field strength (blue circle, yellow square, red diamond, and green pentagon correspond to $\Omega_\mathrm{c}/\gamma_{13}$ = 0.23, 0.32, 0.58, and 0.82, respectively). Dashed and solid lines are the best fit corresponding to $B_\mathrm{EIT}$ and $B_\mathrm{ATS}$ models, respectively. For small $\Omega_\mathrm{c}/\gamma_{13}<0.5$, EIT model (blue and yellow dashed) fits better than ATS model (blue and yellow solid line), while for intermediate value of $\Omega_\mathrm{c}$, both models (red dashed and line) fit well. For strong  $\Omega_\mathrm{c}/\gamma_{13}>0.5$, one obtains a good fit for ATS model (green line), while EIT model (green dashed) fits poorly. From this fit we obtain the relative likelihood ($L_i$) of a model given by Akaike weight $w_i$ as described in section 3, for a particular control field strength. Figure~\ref{fig:eAIC}(b) shows the Akaike weight $\textrm{w}_i$ (hollow square) and per-point weight $\bar{w_i}$ (solid square) with control field strength. Both $w_i$ and $\bar{w}_i$ shows a transition at $\Omega_\mathrm{c}^\mathrm{t}/\gamma_{13}\simeq0.45$, below which EIT model (red) dominates and above ATS model (blue). The transition point obtained from the experimentally measured quantifier $C$ is very close to the AIC test on $B_i$ model ($|\rho_{12}|$) for ideal case (Eq.~\ref{eqn:rho12}). The possible reason of this small deviation is coming from considering negligible ground state decoherence ($\gamma_{12}<<\gamma_{13}$) for numerical computation of $|\rho_{12}|$. However, for a physical system such as thermal ensemble of atoms, the ground state decoherence is finite and can not be ignored. As a result, the transition point for the quantifier is lower than $\Omega_\mathrm{c}^\mathrm{t}/\gamma_{13}=0.5$. While, the test on the quantifier conclusively differentiates between EIT and ATS regimes and reveals the transition point close to the predicted value, the possibility of applying the AIC test to the probe absorption profile is diminished by Doppler effects and power broadening, which wash away any signature of the EIT to ATS transition~\cite{arif18}. This highlights the importance of using coherence when distinguishing between EIT and ATS.

\section{Conclusion}

In conclusion, our study offers a new perspective into the discrimination of EIT and ATS using a robust indicator. We have shown that the ground state coherence $|\rho_{12}|$ gives a distinct transition between EIT and ATS as opposed to the case in Im($\rho_{13}$), when shapes are compared. The transition point obtained from recent studies based on AIC test for Im($\rho_{13}$) shows disagreement with the theoretical threshold value of $\Omega_\mathrm{c}^\mathrm{t}/\gamma_{13}=0.5$~\cite{sanders11,Laurat13,Peng2014,Olga08,Salloum10}, which is obtained by pole analysis of the off-diagonal density element $\rho_{13}$. Shape comparison for probe susceptibility based on this test does not show a clear transition between the two regions because of its similar feature in both the regions. Moreover, probe response also includes incoherent effects which makes the test more ambiguous. Laurat et al. shows that different incoherent effects can take the transition point even further from the threshold point~\cite{Laurat13}. However, $|\rho_{12}|$ is immune to decoherence and has completely distinct feature in the two regimes. One should directly look at ground state coherence instead of probe absorption for unambiguous distinction. We show that AIC weight shows distinct two regions for $|\rho_{12}|$ and the transition point agrees with the theoretical calculation. Furthermore, we apply the test on our reported coherence quantifier which captures the generated ground state coherence for a thermal ensemble of rubidium atoms~\cite{arif18}. AIC weight shows a sharp transition at $\Omega_\mathrm{c}^\mathrm{t}/\gamma_{13}\simeq0.45$, which is nearly equal to the value obtained from theoretical analysis. This 10$\%$ deviation in the transition point for the quantifier $C$ coming from finite ground state decoherence ($\gamma_{12}$), which is taken to be small compared to $\gamma_{13}$ ($\gamma_{12}<<\gamma_{13}$) while applying the AIC test on numerically computed $|\rho_{12}|$ revealing $\Omega_\mathrm{c}^\mathrm{t}/\gamma_{13}=0.5$. In room temperature atoms, where Doppler effects and power broadening make it is almost impossible to apply AIC test and observe any signature of a transition on the probe absorption profile~\cite{arif18}, the coherence quantifier discerns the two regions effectively and the tests reveals the transition point with absolute certainty.

A.W.L and P. A. acknowledge support from UGC. The work was supported by SERB-DST (Grant No. SERB/PHY/2015404).



%

\end{document}